\documentclass[12pt]{article}
\usepackage{amsmath,amssymb,amsfonts, amsbsy, epsfig, subfig}
\usepackage[usenames]{color}
\usepackage{rotating}

\newcommand{\ignore}[1]{}{}

\newtheorem{theorem}{Theorem}[section]
\newtheorem{corollary}{Corollary}[section]
\newtheorem{proposition}{Proposition}[section]
\newtheorem{lemma}{Lemma}[section]
\newtheorem{example}{Example}[section]

\newtheorem{definition}{Definition}[section]

\newcommand{\beq}{\begin{equation}}
\newcommand{\eeq}{\end{equation}}
\newcommand{\beas}{\begin{eqnarray*}}
\newcommand{\eeas}{\end{eqnarray*}}
\newcommand{\bea}{\begin{eqnarray}}
\newcommand{\eea}{\end{eqnarray}}
\newcommand{\bei}{\begin{itemize}}
\newcommand{\eei}{\end{itemize}}
\newcommand{\ben}{\begin{enumerate}}
\newcommand{\een}{\end{enumerate}}
\newcommand{\bet}{\begin{theorem}}
\newcommand{\eet}{\end{theorem}}
\newcommand{\bel}{\begin{lemma}}
\newcommand{\eel}{\end{lemma}}
\newcommand{\bep}{\begin{proposition}}
\newcommand{\eep}{\end{proposition}}
\newcommand{\bed}{\begin{definition}}
\newcommand{\eed}{\end{definition}}
\newcommand{\bec}{\begin{corollary}}
\newcommand{\eec}{\end{corollary}}
\newcommand{\bex}{\begin{example}}
\newcommand{\eex}{\end{example}}
\newcommand{\qed}{\quad\hbox{\vrule width 4pt height 6pt depth 1.5pt}}

\newcommand{\ep}{\epsilon}

\newcommand{\argmin}{\mathop{\rm arg\min}}

\def\0{\boldsymbol{0}}
\def\F{\boldsymbol{F}}

\def\R{\boldsymbol{R}}

\def\A{\boldsymbol{A}}

\def\X{\boldsymbol{X}}

\def\0{\boldsymbol{0}}

\def\X{\boldsymbol{X}}

\def\pr{\textsf{P}} 
\def\ep{\textsf{E}} 
\def\Var{\textsf{Var}} 

\def\liminf{\mathop{\underline{\rm lim}}}

\newcommand{\bc}{b}

\begin{document}

\title{Phase Transition and Regularized Bootstrap in Large Scale $t$-tests with False Discovery Rate Control
}
\author{Weidong Liu\footnote{Department of Mathematics and Institute of Natural Sciences, Shanghai Jiao Tong University. Research supported by NSFC, Grant No.11201298 and No.11322107, the Program for Professor of Special Appointment (Eastern Scholar) at Shanghai Institutions of Higher Learning,  Shanghai Pujiang Program, Foundation for the Author of National Excellent Doctoral Dissertation
of PR China and  Program for New Century Excellent Talents in University. } and Qi-Man Shao\footnote{Department of Statistics, The Chinese University of Hong Kong.
 Research partially supported by Hong Kong RGC GRF 603710 and 403513.
 }}

\maketitle

\begin{abstract}
Applying Benjamini and Hochberg (B-H) method to multiple Student's $t$ tests is a popular technique in gene selection in microarray data analysis. Because of the non-normality of the population, the true p-values of the hypothesis tests are typically unknown. Hence, it is common to use the standard normal distribution $N(0,1)$,  Student's $t$ distribution $t_{n-1}$ or the bootstrap method to
 estimate the p-values. In this paper, we first study $N(0,1)$ and  $t_{n-1}$  calibrations.
 We prove that,  when the population has the finite 4-th moment and the dimension $m$ and the sample size $n$ satisfy $\log m=o(n^{1/3})$, B-H method  controls the false discovery rate (FDR) at a given level $\alpha$ asymptotically with p-values estimated from $N(0,1)$ or $t_{n-1}$ distribution.
  However,  a phase transition  phenomenon occurs  when  $\log m\geq c_{0}n^{1/3}$.
In this case, the FDR of B-H method may be larger than $\alpha$ or even tends to one.
In contrast, the bootstrap calibration  is accurate for $\log m=o(n^{1/2})$ as long as the underlying distribution has the sub-Gaussian tails. However, such light tailed condition can not be weakened in general. The simulation study shows that for the heavy tailed distributions, the bootstrap
 calibration is very conservative. In order to solve this problem, a regularized bootstrap correction is proposed and is shown to be robust to the tails of the distributions.
The simulation study shows that the regularized bootstrap method performs better than the usual bootstrap method.
\end{abstract}

\section{Introduction}

Multiple Student's $t$ tests  often  arise in many real applications such as gene selection. Consider
$m$   tests on the mean values
\begin{eqnarray*}
H_{0i}: ~\mu_{i}=0\quad\mbox{versus\quad} H_{1i}: ~\mu_{i}\neq 0,\quad 1\leq i\leq m.
\end{eqnarray*}
A popular procedure
is using  Benjamini and Hochberg (B-H) method 
 to  search significant findings with the  false discovery rate (FDR) controlled at a given level $0<\alpha<1$, that is,
$$
\ep\Big{[}\frac{\text{V}}{\text{R}\vee 1}\Big{]}\leq \alpha,
$$
where $\text{V}$ is the number of wrongly rejected hypotheses and $\text{R}$ is the total number of rejected hypotheses. The seminal work of Benjamini and Hochberg (1995) is  to reject the null hypotheses for which $p_{i}\leq p_{(\hat{k})}$, where $p_{i}$ is the p-value for $H_{0i}$,
\begin{eqnarray}\label{a0}
\hat{k}=\max\{0 \leq i \leq m:\  p_{(i)}\leq \alpha i/m\},
\end{eqnarray}
and $p_{(1)}\leq\cdots\leq p_{(m)}$ are the order p-values.
Let $T_{1},\ldots,T_{m}$ be Student's $t$ test statistics
\begin{eqnarray*}
T_{i}=\frac{\bar{X}_{i}}{\hat{s}_{ni}/\sqrt{n}},
\end{eqnarray*}
where
\begin{eqnarray*}
\bar{X}_{i}=\frac{1}{n}\sum_{k=1}^{n}X_{ki},\quad \hat{s}^{2}_{ni}=\frac{1}{n-1}\sum_{k=1}^{n}(X_{ki}-\bar{X}_{i})^{2},
\end{eqnarray*}
and $(X_{k1},\ldots,X_{km})^{'}$, $1\leq k\leq n$, are i.i.d. random samples from $(X_{1},\ldots,X_{m})^{'}$. When $T_{1},\ldots,T_{m}$ are independent and the true p-values $p_{i}$ are known, Benjamini and Hochberg (1995)
showed that B-H method controls the FDR at level $\alpha$.

In many applications, the distributions of $X_{i}$, $1\leq i\leq m$, are   non-Gaussian. Hence,
it is impossible to know the exact null distributions of $T_{i}$ and the true p-values. In the application of B-H method, the p-values are actually some estimators.  By the central limit theorem, it is common to use the standard normal distribution $N(0,1)$ or  Student's $t$ distribution $t_{n-1}$ to
 estimate the p-values, where $t_{n-1}$ denotes Student's $t$ random variable with $n-1$ degrees of freedom.  In a microarray analysis, Efron (2004) observed that the choices of null distributions will substantially affect the simultaneous
inference procedure.  However, a systematic theoretical study on the influence of the estimated $p$-values is still lack. It is important to know
how accurate $N(0,1)$ and  $t_{n-1}$ calibrations can be.
  In this paper, we will show that $N(0,1)$ and  $t_{n-1}$  calibrations are accurate when $\log m=o(n^{1/3})$. Under the finite 4th moment of $X_{i}$,
the FDR of B-H method with the estimated p-values $p_{i}=2-2\Phi(T_{i})$ or $p_{i}=2-2\Psi(T_{i})$ will converge to $\alpha m_{0}/m$, where $m_{0}$ is the number of true null hypotheses, $\Phi(t)$ is the standard normal distribution and  $\Psi(t)=\pr(t_{n-1}\leq t)$. However, when $\log m\geq c_{0}n^{1/3}$ for some $c_{0}>0$,
$N(0,1)$ and  $t_{n-1}$  calibrations  may not work well and a phase transition  phenomenon occurs. Under  $\log m\geq c_{0}n^{1/3}$ and the average of skewnesses $\tau=\liminf_{m\rightarrow\infty}m_{0}^{-1}\sum_{i\in\mathcal{H}_{0}}|\ep X_{i}^{3}/\sigma_{i}^{3}|>0$,  we will show that the FDR of B-H method satisfies
$\lim_{(m,n)\rightarrow\infty}FDR\geq \kappa$ for some constant $\kappa>\alpha$, where $\mathcal{H}_{0}=\{i: \mu_{i}=0\}$. Furthermore, if $\log m/n^{1/3}\rightarrow\infty$, then $\lim_{(m,n)\rightarrow\infty}FDR=1$.  This indicates that
$N(0,1)$ and  $t_{n-1}$  calibrations are inaccurate when the average of skewnesses $\tau\neq 0$ in the ultra high dimensional setting.

It is well known that bootstrap  is an effective way to improve the accuracy on   the exact null distribution approximation. Fan, Hall and Yao (2007)
showed that, for the bounded noise, the bootstrap can improve the accuracy and allow higher dimension $\log m=o(n^{1/2})$ on controlling the family-wise error rate. Delaigle, Hall and Jin (2011) showed that the bootstrap method shares significant advantages on higher criticism.
In this paper, we show that,  when the bootstrap calibration is used and $\log m=o(n^{1/2})$,
B-H method can  control FDR at level $\alpha$, i.e. $\lim_{(m,n)\rightarrow\infty}FDR/(\alpha m_{0}/m)=1$. In our results, we assume the sub-Gaussian tails
instead of the  bounded noise in Fan, Hall and Yao (2007).

Although the bootstrap method allows a higher dimension, the light-tailed condition  can not be weakened in general. The simulation study shows that the
bootstrap method is very conservative for the heavy-tailed distributions. To solve this problem, we will propose a regularized bootstrap method which is robust
to the tails of the distributions. The proposed regularized bootstrap only requires the finite 6th moment.
Also, the dimension can be as large as $\log m=o(n^{1/2})$.

It is also not uncommon in real applications that $X_{1},\ldots,X_{m}$ are dependent. This results in the dependency between $T_{1},\ldots,T_{m}$.
 In this paper, we will obtain some similar results for  B-H method under a general weak dependence condition. It should be noted that much work has been done on the robustness of FDR controlling method against dependence. Benjamini and Yekutieli (2001) proved that the B-H procedure  controls FDR under positive regression dependency.
Storey (2003),  Storey, Taylor and Siegmund (2004), Ferreira and Zwinderman (2006) imposed  a dependence  condition  that requires the law of large numbers for the empirical distributions under the null and alternative
hypothesis. Wu (2008) developed a FDR controlling procedures for the data coming from special models such as time series model. However, to satisfy the conditions in most of the existing methods, it is often necessary to assume  the number of true alternative hypotheses $m_{1}$
is asymptotically $\pi_{1}m$ with some $\pi_{1}>0$. They exclude the sparse setting $m_{1}=o(m)$ which is important in applications such as gene selection.
For example, if $m_{1}=o(m)$, then the conditions of Theorem 4 in
Storey, Taylor and Siegmund (2004) and  the conditions of main results in Wu (2008) will be violated.  On the other hand, our results on FDR control under dependence allows $m_{1}\leq \gamma m$ for some $\gamma<1$.

The rest of this paper is organized as follow. In Section 2.1, we will show the robustness  and the phase transition phenomenon for $N(0,1)$ and  $t_{n-1}$
calibrations. In Section 2.2, we show that the bootstrap calibration can improve the FDR control. The regularized bootstrap method is proposed in Section 3. The results are extended to the
dependence case in Section 4. The simulation study is presented in Section 5 and the proofs are given in Section 6.

\section{Main results}

\subsection{Robustness and phase transition}

In this section, we assume  Student's $t$ test statistics $T_{1},\ldots,T_{m}$ are independent. The results will be extended to the dependent case in Section 4.
Before stating the main theorems, we introduce some notations.
Let $\hat{p}_{i,\Phi}=2-2\Phi(|T_{i}|)$
and $\hat{p}_{i,\Psi}=2-2\Psi(|T_{i}|)$ be the $p$-values calculated from the standard normal distribution and the $t$-distribution respectively.
Let FDR$_{\Phi}$ and FDR$_{\Psi}$ be the FDR of B-H method with $\hat{p}_{i,\Phi}$ and $\hat{p}_{i,\Psi}$ in (\ref{a0}) respectively.
Let R be the total number of rejections. The critical values of the tests are then $\hat{t}_{\Phi}=\Phi^{-1}(1-\alpha\text{R}/(2m))$ and $\hat{t}_{\Psi}=\Psi^{-1}(1-\alpha\text{R}/(2m))$.
Set $Y_{i}=(X_{i}-\mu_{i})/\sigma_{i}$ with $\sigma^{2}_{i}=\Var(X_{i})$, $1\leq i\leq m$.

Throughout this paper, we assume  $m_{1}\leq \gamma m$ for some $\gamma<1$, which includes the important sparse setting $m_{1}=o(m)$.

\begin{theorem}\label{th1} Suppose $X_{1},\ldots,X_{m}$ are independent and $\log m=o(n^{1/2})$. Assume that $\max_{1\leq i\leq m}\ep Y_{i}^{4}\leq \bc_{0}$ for some constant $\bc_{0}>0$ and
\begin{eqnarray}\label{c1}
Card\Big{\{}i: |\mu_{i}/\sigma_{i}|\geq 4\sqrt{\log m/n}\Big{\}}\rightarrow\infty.
\end{eqnarray}
Then
\begin{eqnarray*}
\lim_{(n,m)\rightarrow\infty} \frac{FDR_{\Phi}}{\frac{m_{0}}{m}\alpha\kappa_{\Phi}}=1\mbox{\quad and\quad} \lim_{(n,m)\rightarrow\infty} \frac{FDR_{\Psi}}{\frac{m_{0}}{m}\alpha\kappa_{\Psi}}=1,
\end{eqnarray*}
where
\begin{eqnarray*}
\kappa_{\Phi}&=&\ep[ \hat{\kappa}_{\Phi} I\{ \hat{\kappa}_{\Phi}\leq 2(\alpha-\alpha\gamma)^{-1}\}],\cr
\hat{\kappa}_{\Phi}&=&\frac{\sum_{i\in\mathcal{H}_{0}}\Big{\{}\exp\Big{(}\frac{\hat{t}^{3}_{\Phi}\ep X_{i}^{3}}{\sqrt{n}\sigma_{i}^{3}}\Big{)}+\exp\Big{(}-\frac{\hat{t}^{3}_{\Phi}\ep X_{i}^{3}}{\sqrt{n}\sigma_{i}^{3}}\Big{)}\Big{\}}}{2m_{0}}
\end{eqnarray*}
satisfying $1+o(1)\leq \kappa_{\Phi}\leq m/(\alpha m_{0})+o(1)$,
and $\kappa_{\Psi}$ is defined in the same way.
\end{theorem}

Recall that $\tau=\liminf_{m\rightarrow\infty}m_{0}^{-1}\sum_{i\in\mathcal{H}_{0}}|\ep Y_{i}^{3}|$. We have the following corollary.
\begin{corollary}\label{co1} Assume the conditions in Theorem \ref{th1} hold.

\begin{itemize}

\item[(i).] Under $\log m=o(n^{1/3})$, we have
$\lim_{(n,m)\rightarrow\infty} FDR_{\Phi}/(\alpha m_{0}/m)=1$.

\item[(ii).] Suppose $\log m\geq c_{0}n^{1/3}$ for some $c_{0}>0$ and
$m_{1}=\exp(o(n^{1/3}))$. Assume that $\tau>0$. We have $\liminf_{(n,m)\rightarrow\infty} FDR_{\Phi}\geq \beta$ for some constant $\beta>\alpha$.

\item[(iii).]  Suppose $\log m/n^{1/3}\rightarrow\infty$ and
$m_{1}=\exp(o(n^{1/3}))$. Assume that $\tau>0$. We have $\lim_{(n,m)\rightarrow\infty} FDR_{\Phi}=1$.
\end{itemize}

The same conclusions hold for $FDR_{\Psi}$.
\end{corollary}

Theorem \ref{th1} and Corollary \ref{co1} show that, when $\log m=o(n^{1/3})$, $N(0,1)$ and  $t_{n-1}$ calibrations are accurate. Note that only a finite fourth moment of $Y_{i}$ is required. Furthermore, if the skewnesses $\ep Y_{i}^{3}=0$ for $i\in\mathcal{H}_{0}$, then the dimension can be as large as $\log m=o(n^{1/2})$. However, a phase transition occurs if the average of skewnesses  $\tau>0$, for example, for the exponential distribution.
The FDR of B-H method  will be greater than $\alpha$ as long as $\log m\geq c_{0}n^{1/3}$ and will converge to one when $\log m/n^{1/3}\rightarrow\infty$.

Corollary 2.1 also indicates that, in the study of large scale testing problem, the choice of asymptotic null distributions is important. When the dimension is much larger than
the sample size, an inadequate choice such as $N(0,1)$ may result in a  high FDR. This will be further verified by our simulation study in Section 5. Hence, in the problems on large scale tests, assuming the true p-values are known  may be over-idealistic.

\subsection{Bootstrap  calibration}

In this section, we show that the bootstrap  procedure can improve the accuracy on the control of FDR.
Write $\mathcal{X}_{i}=\{X_{1i},\ldots,X_{ni}\}$. Let $\mathcal{X}^{*}_{ki}=\{X^{*}_{1ki},\ldots,X^{*}_{nki}\}$, $1\leq k\leq N$,
be resamples
drawn randomly with replacement from $\mathcal{X}_{i}$. Let $T^{*}_{ki}$ be  Student's $t$ test statistics constructed from  $\{X^{*}_{1ki}-\bar{X}_{i},\ldots,X^{*}_{nki}-\bar{X}_{i}\}$. We use
$G^{*}_{N,m}(t)=\frac{1}{Nm}\sum_{k=1}^{N}\sum_{i=1}^{m}I\{|T^{*}_{ki}|\geq t\}$ to approximate the null distribution and define the $p$-values by
$\hat{p}_{i,B}=G^{*}_{N,m}(|T_{i}|)$. Let FDR$_{B}$ denote the FDR of B-H method with $\hat{p}_{i,B}$ in (\ref{a0}).

\begin{theorem}\label{th2-2} Suppose that $\max_{1\leq i\leq m}\ep e^{tY_{i}^{2}}\leq K$ for some constants $t>0$ and $K>0$  and the conditions in Theorem \ref{th1} hold.

\begin{itemize}

\item[(1).] Under $\log m=o(n^{1/3})$, we have
$\lim_{(n,m)\rightarrow\infty} FDR_{B}/(\alpha m_{0}/m)=1$.

\item[(2).] If  $\log m=o(n^{1/2})$ and $m_{1}\leq m^{\eta}$ for some $\eta<1$, then $\lim_{(n,m)\rightarrow\infty} FDR_{B}/(\alpha m_{0}/m)=1$.
\end{itemize}

\end{theorem}

Another common bootstrap method is to estimate the $p$-values individually by $\breve{p}_{i,B}=G^{*}_{i}(T_{i})$, where
$G^{*}_{i}(t)=\frac{1}{N}\sum_{k=1}^{N}I\{T^{*}_{ki}\geq t\}$; see Fan, Hall and Yao (2007) and Delaigle, Hall and Jin (2011). Similar results as Theorem
\ref{th2-2} can be obtained if $N$ is large enough (e.g. $N\geq m$). Note that in Theorem \ref{th2-2}, $N\geq 1$ is sufficient because we use
the average of all $m$ variables.

Fan, Hall and Yao (2007) proved that, the bootstrap
calibration is accurate for the control of family-wise error rate if $\log m=o(n^{1/2})$ and $\pr(|Y_{i}|\leq C)=1$ for $1\leq i\leq m$.
Our result on FDR control only requires the sub-Gaussian tails which is weaker than the bounded noise.

\noindent{\bf Remark.}
{\em The light-tailed moment condition for bootstrap calibration.}
The bootstrap method has often been used in multiple Student's $t$ tests in real applications.
Fan, Hall and Yao (2007) and Delaigle, Hall and Jin (2011) have proved that the bootstrap method provides a more accurate p-values than the normal or $t_{n-1}$ approximation for the light-tailed distributions.
Theorem 2.2 shows that the bootstrap method allows a higher dimension $\log m=o(n^{1/2})$ for FDR control
when $\max_{1\leq i\leq m}\ep e^{tY_{i}^{2}}\leq K$.
However,
it is not necessary that the real data would satisfy such light tailed condition.
We  argue that the light tailed condition can not be weakened
in general when the bootstrap method is used. Denote the conditional tails of distribution of the bootstrap version for Student's $t$ statistic by $G^{*}_{i}(t)=\pr(|T_{i}^{*}|\geq t|\mathcal{X})$, where $\mathcal{X}=\{\mathcal{X}_{1},\ldots,\mathcal{X}_{m}\}$. Gin\'{e}, et al. (1997) proved that Student's $t$ statistic is asymptotically normal if and only if the underlying distribution of
  the population is in the domain of
attraction of the normal law. This implies  any $\alpha$-th moment ($0<\alpha<2$) of the underlying distribution  is finite. Hence, to ensure $G^{*}_{i}(t)\rightarrow 2-2\Phi(t)$,
we often need
\begin{eqnarray}\label{a51}
\max_{1\leq i\leq m}\frac{1}{n}\sum_{k=1}^{n}|X_{ki}-\bar{X}_{i}|^{\alpha}\leq K
\end{eqnarray}
 for any $0<\alpha<2$. Suppose the components $X_{1},\ldots,X_{m}$ are independent and identically distributed and $\log m\asymp n^{\gamma}$,
 $\gamma>0$. A necessary condition for (\ref{a51}) is $\ep \exp(t_{0}|X_{1}|^{\alpha \gamma})<\infty$ for some $t_{0}>0$. So when $\log m=o(n^{1/3})$,
 the bootstrap method requires a much more stringent moment condition than $N(0,1)$ or $t_{n-1}$  calibration.
 From the above analysis, we can see that the bootstrap calibration may not always outperform the $N(0,1)$ or $t_{n-1}$  calibration. In particular, when
 the distribution is symmetric, $N(0,1)$ and $t_{n-1}$  approximations can even perform better than the bootstrap method.
 This will be
further verified by the simulation study in Section 5.

\section{Regularized bootstrap in large scale tests}

In this section, we introduce a regularized bootstrap method that is robust for heavy-tailed distributions and
the dimension $m$ can be as large as $e^{o(n^{1/2})}$.
For the regularized bootstrap method, the finite 6th moment condition is enough.  Let $\lambda_{ni}\rightarrow\infty$ be a regularized parameter. Define
\begin{eqnarray*}
\hat{X}_{ki}=X_{ki}I\{|X_{ki}|\leq \lambda_{ni}\}, \quad 1\leq k\leq n,\quad 1\leq i\leq m.
\end{eqnarray*}
Write $\hat{\mathcal{X}}_{i}=\{\hat{X}_{1i},\ldots,\hat{X}_{ni}\}$. Let $\hat{\mathcal{X}}^{*}_{ki}=\{\hat{X}^{*}_{1ki},\ldots,\hat{X}^{*}_{nki}\}$, $1\leq k\leq N$,
be resamples
drawn independently and uniformly with replacement from $\hat{\mathcal{X}}_{i}$. Let $\hat{T}^{*}_{ki}$ be  Student's $t$ test statistics constructed from  $\{\hat{X}^{*}_{1ki}-\hat{X}_{i},\ldots,\hat{X}^{*}_{nki}-\hat{X}_{i}\}$, where $\hat{X}_{i}=\frac{1}{n}\sum_{k=1}^{n}\hat{X}_{ki}$. We use
$\hat{G}^{*}(t)=\frac{1}{Nm}\sum_{k=1}^{N}\sum_{i=1}^{m}I\{|\hat{T}^{*}_{ki}|\geq t\}$ to approximate the null distribution and define the $p$-values by
$\hat{p}_{i,RB}=\hat{G}^{*}(|T_{i}|)$. Let FDR$_{RB}$  be the FDR of B-H method with $\hat{p}_{i,RB}$ in (\ref{a0}).

\begin{theorem}\label{th2-222} Assume that $\max_{1\leq i\leq m}\ep X_{i}^{6}\leq K$ for some constant $K>0$. Suppose $X_{1},\ldots,X_{m}$ are independent, (\ref{c1}) holds and $\min_{1\leq i\leq m}\sigma_{ii}\geq c_{0}$ for some $c_{0}>0$. Let $c_{1}(n/\log m)^{1/6}\leq \lambda_{ni}\leq c_{2}(n/\log m)^{1/6}$
for some $c_{1},c_{2}>0$.

\begin{itemize}

\item[(1).] Under $\log m=o(n^{1/3})$, we have
$\lim_{(n,m)\rightarrow\infty} FDR_{RB}/(\alpha m_{0}/m)=1$.

\item[(2).] If  $\log m=o(n^{1/2})$ and $m_{1}\leq m^{\eta}$ for some $\eta<1$, then $\lim_{(n,m)\rightarrow\infty} FDR_{RB}/(\alpha m_{0}/m)=1$.
\end{itemize}

\end{theorem}

In Theorem \ref{th2-222}, we only require $\max_{1\leq i\leq m}\ep X_{i}^{6}\leq K$, which is much weaker than the moment condition in Theorem 2.2.

In the regularized bootstrap method, we need to choose the regularized parameter $\lambda_{ni}$. By Theorem 1.2 in Wang (2005),
equation (2.2) in Shao (1999) and the proof of Theorem 3.1, we have
\begin{eqnarray*}
\pr(|\hat{T}^{*}_{ki}|\geq t|\hat{\mathcal{X}})=\frac{1}{2}G(t)\Big{[}\exp\Big{(}\frac{t^{3}}{\sqrt{n}}\hat{\kappa}_{i}(\lambda_{ni})\Big{)}
+\exp\Big{(}-\frac{t^{3}}{\sqrt{n}}\hat{\kappa}_{i}(\lambda_{ni})\Big{)}\Big{]}(1+o_{\pr}(1)),
\end{eqnarray*}
uniformly for $0\leq t\leq o(n^{1/4})$, where $\hat{\mathcal{X}}=\{\hat{\mathcal{X}}_{1},\ldots,\hat{\mathcal{X}}_{m}\}$,
\begin{eqnarray}\label{a00}
\hat{\kappa}_{i}(\lambda_{ni})=\frac{1}{n\hat{\sigma}_{i}^{3}}\sum_{k=1}^{n}(\hat{X}_{ki}-\hat{X}_{i})^{3}\quad\mbox{and\quad}
\hat{\sigma}^{2}_{i}=\frac{1}{n}\sum_{k=1}^{n}(\hat{X}_{ki}-\hat{X}_{i})^{2}.
\end{eqnarray}
Also,
\begin{eqnarray*}
\pr(|T_{i}|\geq t)=\frac{1}{2}G(t)\Big{[}\exp\Big{(}\frac{t^{3}}{\sqrt{n}}\kappa_{i}\Big{)}
+\exp\Big{(}-\frac{t^{3}}{\sqrt{n}}\kappa_{i}\Big{)}\Big{]}(1+o(1)),
\end{eqnarray*}
uniformly for $0\leq t\leq o(n^{1/4})$, where $\kappa_{i}=\ep Y_{i}^{3}$. A good choice of $\lambda_{ni}$ is to make $\hat{\kappa}_{i}(\lambda_{ni})$ get  close to $\kappa_{i}$.
As $\kappa_{i}$ is unknown, we propose the following cross-validation method.\vspace{3mm}

\noindent{\bf Data-driven choice of $\lambda_{ni}$.} We propose to choose  $\hat{\lambda}_{ni}=|\bar{X}_{i}|+\hat{s}_{ni}\lambda$, where $\lambda$
 will be selected as follow. Split the samples into two parts $\mathcal{I}_{0}=\{1,\ldots, n_{1}\}$ and
 $\mathcal{I}_{1}=\{n_{1}+1,\ldots, n\}$ with  sizes $n_{0}=[n/2]$
and $n_{1}=n-n_{0}$ respectively. For $\mathcal{I}=\mathcal{I}_{0}$ or $\mathcal{I}_{1}$,
let
\begin{eqnarray*}
\hat{\kappa}_{i,\mathcal{I}}=\frac{1}{|\mathcal{I}|\hat{s}_{ni,\mathcal{I}}^{3}}\sum_{k\in\mathcal{I}}(X_{ki}-\bar{X}_{i,\mathcal{I}})^{3},\quad
\hat{s}^{2}_{ni,\mathcal{I}}=\frac{1}{|\mathcal{I}|}\sum_{k\in\mathcal{I}}(X_{ki}-\bar{X}_{i,\mathcal{I}})^{2},\quad
\bar{X}_{i,\mathcal{I}}=\frac{1}{|\mathcal{I}|}\sum_{k\in\mathcal{I}}X_{ki}.
\end{eqnarray*}
Let $\hat{\kappa}_{i,\mathcal{I}}(\lambda_{ni})$, with $\lambda_{ni}=|\bar{X}_{i,\mathcal{I}}|+\hat{s}_{ni,\mathcal{I}}\lambda$, be defined as in (\ref{a00}) based on $\{\hat{X}_{ki}, k\in\mathcal{I}\}$.
Define the risk
\begin{eqnarray*}
R_{j}(\lambda)=\sum_{i=1}^{m}(\hat{\kappa}_{i,\mathcal{I}_{j}}(\lambda_{ni})-\hat{\kappa}_{i,\mathcal{I}_{1-j}})^{2}.
\end{eqnarray*}
 We choose  $\lambda$
by
\begin{eqnarray}\label{a000}
\hat{\lambda}=\argmin_{0<\lambda<\infty}\{R_{0}(\lambda)+R_{1}(\lambda)\}.
\end{eqnarray}
The final regularized parameter is $\hat{\lambda}_{ni}=|\bar{X}_{i}|+\hat{s}_{ni}\hat{\lambda}$.

It is important to investigate the theoretical property of $\hat{\lambda}_{ni}$ and to see whether Theorem 3.1 still hold when $\hat{\lambda}_{ni}$ is used.
We leave this as a future work.

\section{FDR control under dependence}

 To generalize the results to the dependent case, we introduce a class of correlation matrices. Let  $\A=(a_{ij})$ be a symmetric matrix.
 Let $k_m$ and $s_m$  be positive numbers.
  Assume that for
every $1\leq j\leq m$,
\begin{eqnarray}\label{a50}
\text{Card}\{1\leq i\leq m: |a_{ij}|\geq k_{m}\}\leq s_{m}.
\end{eqnarray}
Let $\mathcal{A}(k_{m},s_{m})$ be the class of symmetric matrices satisfying (\ref{a50}). Let $\R=(r_{ij})$  be the correlation matrix of $\X$.
We introduce the following two conditions.

\begin{itemize}

\item[{\bf (C1).}] Suppose that $\max_{1\leq j<j\leq m}|r_{ij}|\leq r$ for some $0<r<1$ and $\R\in \mathcal{A}(k_{m},s_{m})$ with $k_{m}=(\log m)^{-2-\delta}$ and $s_{m}=O(m^{\rho})$ for some $\delta>0$ and $0<\rho<(1-r)/(1+r)$.

\item[{\bf (C1$^{*}$).}] Suppose that $\max_{1\leq j<j\leq p}|r_{ij}|\leq r$ for some $0<r<1$. For each $X_{i}$, assume the number  of variables $X_{j}$ which are dependent with $X_{i}$ is no more than $s_{m}$.\vspace{3mm}
\end{itemize}

(C1) and (C1$^{*}$) impose the weak dependence between $X_{1},\ldots,X_{m}$. In (C1), each variable can be highly correlated with other $s_{m}$ variables and weakly
correlated with the remaining variables. (C1$^{*}$) is stronger than (C1). For each $X_{i}$, (C1$^{*}$) requires the independence between $X_{i}$
and other $m-s_{m}$ variables.

Recall that $m_{1}\leq \gamma m$ for some $\gamma<1$.

\begin{theorem}\label{th21} Assume that $\max_{1\leq i\leq m}\ep Y_{i}^{4}\leq \bc_{0}$ for some constant $\bc_{0}>0$ and (\ref{c1}) holds.

\begin{itemize}
\item[(i).] Under $\log m=O(n^{\zeta})$ for some $0<\zeta<3/23$ and (C1), we have
\begin{eqnarray}\label{th3}
\lim_{(n,m)\rightarrow\infty} \frac{FDR_{\Phi}}{\frac{m_{0}}{m}\alpha}=1,\quad \lim_{(n,m)\rightarrow\infty} \frac{FDR_{\Psi}}{\frac{m_{0}}{m}\alpha}=1
\end{eqnarray}

\item[(ii).] Under $\log m=o(n^{1/3})$ and (C1$^{*}$), we have (\ref{th3}) holds.

\end{itemize}

\end{theorem}

For the bootstrap and regularized procedures, we have the similar results.

\begin{theorem}\label{th22}Suppose that $\max_{1\leq i\leq m}\ep e^{tY_{i}^{2}}\leq K$ and (\ref{c1}) holds.

\begin{itemize}

\item[(1).] Under the conditions of (i) or (ii) in Theorem \ref{th21}, we have $\lim_{(n,m)\rightarrow\infty} \frac{FDR_{B}}{\frac{m_{0}}{m}\alpha}=1$

\item[(2).]  Under (C1$^{*}$), $\log m=o(n^{1/2})$ and $m_{1}\leq m^{\eta}$ for some $\eta<1$, we have
    $$\lim_{(n,m)\rightarrow\infty} \frac{FDR_{B}}{\frac{m_{0}}{m}\alpha}=1 . $$

\end{itemize}

\end{theorem}

\begin{theorem}\label{th222}Suppose that $\max_{1\leq i\leq m}\ep X_{i}^{6}\leq K$ for some constant $K>0$, $\min_{1\leq i\leq m}\sigma_{ii}\geq c_{0}$ for some $c_{0}>0$ and (\ref{c1}) holds.  Let $c_{1}(n/\log m)^{1/6}\leq \lambda_{ni}\leq c_{2}(n/\log m)^{1/6}$
for some $c_{1},c_{2}>0$.

\begin{itemize}

\item[(1).] Under the conditions of (i) or (ii) in Theorem \ref{th21}, we have $\lim_{(n,m)\rightarrow\infty} \frac{FDR_{RB}}{\frac{m_{0}}{m}\alpha}=1$

\item[(2).]  Under (C1$^{*}$), $\log m=o(n^{1/2})$ and $m_{1}\leq m^{\eta}$ for some $\eta<1$, we have
    $$\lim_{(n,m)\rightarrow\infty} \frac{FDR_{RB}}{\frac{m_{0}}{m}\alpha}=1 . $$

\end{itemize}

\end{theorem}

Theorems \ref{th21}-\ref{th222} imply that B-H method remains valid asymptotically
for weak dependence. As the phase transition phenomenon
caused by the growth of the dimension, it would be interesting to investigate when will B-H method fail to control the FDR as the correlation becomes strong.

\section{Numerical Study}

In this section, we first carry out a small simulation to verify the phase transition phenomenon.
Let
\begin{eqnarray}\label{a5-0}
X_{i}=\mu_{i}+(\varepsilon_{i}-\ep \varepsilon_{i}),\quad 1\leq i\leq m,
\end{eqnarray}
 where $(\varepsilon_{1},\ldots,\varepsilon_{m})^{'}$ are i.i.d. random variables. We consider two models for $\varepsilon_{i}$ and $\mu_{i}$.\vspace{2mm}

  {\bf Model 1.} $\varepsilon_{i}$ is the exponential random variable with parameter 1. Let $\mu_{i}=2\sigma\sqrt{\log m/n}$ for $1\leq i\leq m_{1}$
with $m_{1}=0.05 m$ and $\mu_{i}=0$ for $m_{1}<i\leq m$, where $\sigma^{2}=\Var(\varepsilon_{i})$.

    {\bf Model 2.} $\varepsilon_{i}$ is the Gamma random variable with parameter (0.5,1). Let $\mu_{i}=4\sigma\sqrt{\log m/n}$ for $1\leq i\leq m_{1}$
with $m_{1}=0.05 m$ and $\mu_{i}=0$ for $m_{1}<i\leq m$.\vspace{2mm}

In both models, the average of skewnesses $\tau>0$.
We generate $n=30, \ 50$ independent random samples from (\ref{a5-0}).
In our simulation, $\alpha$ is taken to be $0.1,0.2,0.3$ and $m$ is taken to be $500$, $1000$, $3000$. In the usual  bootstrap approximation  and
 the regularized bootstrap approximation, the resampling time $N$ is taken to be 200. The simulation is replicated 500 times and the
empirical FDR and power are summarized in Tables 1 and 2. The empirical power is defined by the average  ratio between the number of correct rejections and
$m_{1}$.
As we can see, due to the nonzero skewnesses  and $m\gg\exp(n^{1/3})$, the empirical FDR$_{\Phi}$  and
FDR$_{\Psi}$ are much larger than the target FDR. The bootstrap method and the regularized bootstrap method provide  more accurate approximations for the true p-values. So the empirical
FDR$_{B}$  and FDR$_{RB}$ are much closer to $\alpha$ than FDR$_{\Phi}$  and
FDR$_{\Psi}$ do. For Models 1 and 2, the bootstrap method and the proposed regularized bootstrap method perform quite similarly.  All of four methods perform better as the sample size $n$ grows from 30 to 50, although the empirical FDR$_{\Phi}$  and
FDR$_{\Psi}$ still have seriously departure from $\alpha$.

Next, we consider the following two models to compare the performance between the four methods when the distributions are symmetric and heavy tailed.

 {\bf Model 3.} $\varepsilon_{i}$ is Student's $t$ distribution with 4 degrees of freedom. Let $\mu_{i}=2\sqrt{\log m/n}$ for $1\leq i\leq m_{1}$
with $m_{1}=0.1 m$ and $\mu_{i}=0$ for $m_{1}<i\leq m$.

 {\bf Model 4.} $\varepsilon_{i}=\varepsilon_{i1}-\varepsilon_{i2}$, where $\varepsilon_{i1}$ and $\varepsilon_{i1}$ are independent lognormal random variables with parameters $(0,1)$. Let $\mu_{i}=4\sqrt{\log m/n}$ for $1\leq i\leq m_{1}$
with $m_{1}=0.1 m$ and $\mu_{i}=0$ for $m_{1}<i\leq m$.

For these two models, the normal approximation performs the best on the control of FDR; see Tables 3 and 4. FDR$_{B}$ is much
smaller than $\alpha$ so the  bootstrap method is quite conservative. This is mainly due to the heavy tails of the $t(4)$ and lognormal distributions.
The regularized bootstrap method works much better than the bootstrap method on the FDR control. From Table 4, we see that it also has the higher powers (power$_{RB}$) than the bootstrap method (power$_{B}$).
Hence, the proposed regularized bootstrap is more robust than the commonly used bootstrap method.

\begin{table}[hptb]\small\addtolength{\tabcolsep}{-4pt}
  \begin{center}
    \caption{Comparison of FDR (FDR=$\alpha$)}
    \begin{tabular}{|c c| @{\hspace{2em}}c@{\hspace{1em}}c@{\hspace{1em}}c|@{\hspace{1em}}c@{\hspace{1em}}c@{\hspace{1em}} c| }
      \hline
      & \multicolumn{4}{c}{$n=30$}&\multicolumn{3}{c|}{$n=50$} \\
      \hline
  $m$  & $ |\quad { \alpha}$ &  0.1 &0.2& 0.3& 0.1&
      0.2 & 0.3\\
      \hline
      & \multicolumn{6}{c}{Exp(1)}&\multicolumn{1}{c|}{} \\
      \hline
       500 & FDR$_{\Phi}$    &  0.3746   &  0.4670         & 0.5422       & 0.2898    &0.3913    & 0.4738   \\
          &FDR$_{\Psi}$   &   0.3081   &  0.4085         & 0.4863       & 0.2482    &0.3501     & 0.4357 \\
          &FDR$_{B}$    &   0.0649   &  0.1730         & 0.2778       & 0.0912    &0.1869     & 0.2845 \\
            &FDR$_{RB}$    &   0.0675   & 0.1761         & 0.2860       & 0.0885    &0.1877    & 0.2851 \\
          \hline
      1000 & FDR$_{\Phi}$    &   0.3762   &  0.4717         & 0.5461       & 0.2916    &0.3962     & 0.4810   \\
          &FDR$_{\Psi}$  &   0.3097  &  0.4113        & 0.4919       & 0.2488    &0.3561     & 0.4404 \\
          &FDR$_{B}$    &   0.0695   & 0.1771         & 0.2860       & 0.0916    &0.1934     & 0.2906 \\
           &FDR$_{RB}$    &   0.0675   & 0.1765         &0.2864      & 0.0919    &0.1921    & 0.2909 \\
          \hline
      3000 & FDR$_{\Phi}$  &   0.3811   &  0.4785         & 0.5517       & 0.2944    &0.3987     & 0.4818   \\
          &FDR$_{\Psi}$   &   0.3129   &  0.4178         & 0.4978       & 0.2510    &0.3580     & 0.4432 \\
          &FDR$_{B}$      &   0.0703   &  0.1810         & 0.2865       & 0.0931    &0.1942     & 0.2938 \\
          &FDR$_{RB}$    &  0.0692   & 0.1775         & 0.2850       & 0.0936    &0.1928    & 0.2922 \\
           \hline
      & \multicolumn{6}{c}{Gamma(0.5,1)}&\multicolumn{1}{c|}{} \\
      \hline
       500 & FDR$_{\Phi}$   &   0.4973   &  0.5780         & 0.6339       &0.3963    &0.4903     & 0.5601   \\
          &FDR$_{\Psi}$   &   0.4436   &  0.5333         & 0.5981      & 0.3593    &0.4567     & 0.5301 \\
          &FDR$_{B}$    &   0.0738   &  0.1751         & 0.2827       & 0.0842    &0.1853     & 0.2939 \\
          &FDR$_{RB}$    &   0.0755   & 0.1758        & 0.2943       & 0.0883    &0.1882    & 0.2941 \\
          \hline
      1000 & FDR$_{\Phi}$   &   0.5019   &  0.5810         & 0.6368       & 0.3992    &0.4929     & 0.5617   \\
          &FDR$_{\Psi}$    &   0.4480   &  0.5382         & 0.6019       & 0.3624    &0.4605     & 0.5322 \\
          &FDR$_{B}$  &   0.0753   &  0.1758         & 0.2867       & 0.0879    &0.1883     & 0.2932 \\
          &FDR$_{RB}$    &   0.0688   & 0.1740         & 0.2823       & 0.0859    &0.1902    & 0.2926 \\
          \hline
      3000 & FDR$_{\Phi}$  &   0.5025   &  0.5813         & 0.6375       & 0.4023    &0.4952     & 0.5634   \\
          &FDR$_{\Psi}$    &   0.4483   &  0.5386         & 0.6021       & 0.3647    &0.4636     & 0.5351 \\
          &FDR$_{B}$       &   0.0737   &  0.1769         & 0.2873       & 0.0864    &0.1909     & 0.2948 \\
          &FDR$_{RB}$    &  0.0723    & 0.1741         & 0.2847       & 0.0854    &0.1878    & 0.2911 \\
          \hline
    \end{tabular}

    \label{tb:simu3}
  \end{center}
\end{table}

\begin{table}[hptb]\small\addtolength{\tabcolsep}{-4pt}
  \begin{center}
    \caption{Comparison of power (FDR=$\alpha$)}
    \begin{tabular}{|c c| @{\hspace{2em}}c@{\hspace{1em}}c@{\hspace{1em}}c|@{\hspace{1em}}c@{\hspace{1em}}c@{\hspace{1em}} c| }
      \hline
      & \multicolumn{4}{c}{$n=30$}&\multicolumn{3}{c|}{$n=50$} \\
      \hline
  $m$  & $ |\quad { \alpha}$ &  0.1 &0.2& 0.3& 0.1&
      0.2 & 0.3\\
      \hline
      & \multicolumn{6}{c}{exp(1)}&\multicolumn{1}{c|}{} \\
      \hline
       500 & power$_{\Phi}$    & 0.9998   &  1.0000          & 1.0000        & 0.9999     &1.0000     & 1.0000    \\
          &power$_{\Psi}$   &   0.9995    &  0.9998          & 1.0000        & 0.9997     &1.0000     & 1.0000  \\
          &power$_{B}$    &   0.7473    &  0.9852          & 0.9990       & 0.9831     &0.9986     & 0.9998  \\
          &power$_{RB}$    &   0.7371    & 0.9848         & 0.9989       & 0.9839     &0.9981      & 0.9994  \\
          \hline
      1000 & power$_{\Phi}$    &   0.9999   &  1.0000        & 1.0000        & 1.0000     &1.0000      & 1.0000    \\
          &power$_{\Psi}$  &   0.9997   &  1.0000         & 1.0000        & 1.0000    &1.0000      & 1.0000  \\
          &power$_{B}$    &   0.8873   & 0.9943          & 0.9991        & 0.9945     &0.9998      & 1.0000  \\
          &power$_{RB}$    &   0.8880   & 0.9936         & 0.9995       & 0.9942    &0.9996    & 0.9999 \\
          \hline
      3000 & power$_{\Phi}$  &   1.0000    & 1.0000          & 1.0000         & 1.0000     &1.0000      & 1.0000   \\
          &power$_{\Psi}$   &   0.9999    &  1.0000         & 1.0000         & 1.0000     &1.0000      & 1.0000  \\
          &power$_{B}$      &   0.9642    &  0.9984          & 0.9999        & 0.9987     &1.0000    & 1.0000  \\
          &power$_{RB}$    &   0.9650   & 0.9983         & 0.9999       & 0.9989    &0.9999    &1.0000 \\
           \hline
      & \multicolumn{6}{c}{Gamma(0.5,1)}&\multicolumn{1}{c|}{} \\
      \hline
       500 & power$_{\Phi}$   &  1.0000    &  1.0000          & 1.0000        &1.0000     &1.0000      & 1.0000   \\
          &power$_{\Psi}$   &   1.0000    &  1.0000         & 1.0000     &1.0000     &1.0000      & 1.0000  \\
          &power$_{B}$    &   0.9986    &  0.9999         & 1.0000       &   1.0000    &  1.0000         & 1.0000 \\
          &power$_{RB}$    &   0.9982   & 0.9950         & 0.9994       & 1.0000    &1.0000    & 1.0000 \\
          \hline
      1000 & power$_{\Phi}$   &  1.0000     &  1.0000         & 1.0000       &1.0000     &1.0000      & 1.0000   \\
          &power$_{\Psi}$    & 1.0000    &  1.0000          & 1.0000       &1.0000     &1.0000      & 1.0000  \\
          &power$_{B}$  &   0.9988    & 1.0000      & 1.0000        &   1.0000    &  1.0000         & 1.0000  \\
          &power$_{RB}$    &   0.9584   &0.9978         & 0.9998       & 1.0000    &1.0000    & 1.0000 \\
          \hline
      3000 & power$_{\Phi}$  &   1.0000    &  1.0000         & 1.0000        &1.0000     &1.0000      & 1.0000   \\
          &power$_{\Psi}$    &   1.0000   & 1.0000         & 1.0000       &1.0000     &1.0000      & 1.0000  \\
          &power$_{B}$       &   0.9994    & 1.0000        & 1.0000        &   1.0000    &  1.0000         & 1.0000 \\
          &power$_{RB}$    &   0.9822   & 0.9988         & 0.9999       & 1.0000    &1.0000    & 1.0000 \\
          \hline
    \end{tabular}

    \label{tb:simu3}
  \end{center}
\end{table}

\begin{table}[hptb]\small\addtolength{\tabcolsep}{-4pt}
  \begin{center}
    \caption{Comparison of FDR (FDR=$\alpha$)}
    \begin{tabular}{|c c| @{\hspace{2em}}c@{\hspace{1em}}c@{\hspace{1em}}c|@{\hspace{1em}}c@{\hspace{1em}}c@{\hspace{1em}} c| }
      \hline
      & \multicolumn{4}{c}{$n=30$}&\multicolumn{3}{c|}{$n=50$} \\
      \hline
  $m$  & $ |\quad { \alpha}$ &  0.1 &0.2& 0.3& 0.1&
      0.2 & 0.3\\
      \hline
      & \multicolumn{6}{c}{t(4)}&\multicolumn{1}{c|}{} \\
      \hline
       500 & FDR$_{\Phi}$    &  0.1147   &  0.2129          & 0.3082        & 0.1006     &0.1958     & 0.2900    \\
          &FDR$_{\Psi}$   &   0.0704    &  0.1536          & 0.2442        & 0.0741     &0.1600      & 0.2514  \\
          &FDR$_{B}$    &   0.0358   &  0.1112          & 0.1991       & 0.0438     &0.1214      & 0.2022  \\
           &FDR$_{RB}$    &   0.0612   & 0.1435          &0.2348        & 0.0693      &0.1565      & 0.2448  \\
          \hline
      1000 & FDR$_{\Phi}$    &   0.1170   &  0.2153         & 0.3083        & 0.1014     &0.1968      & 0.2905    \\
          &FDR$_{\Psi}$  &   0.0705   &  0.1571         & 0.2472        & 0.0756    &0.1618      & 0.2532  \\
          &FDR$_{B}$    &   0.0341    & 0.1072         & 0.1904       & 0.0511     &0.1333      & 0.2242  \\
               &FDR$_{RB}$    &   0.0593   &  0.1432          & 0.2324        & 0.0718      &0.1584      & 0.2507  \\
          \hline
      3000 & FDR$_{\Phi}$  &   0.1166    &  0.2150          & 0.3093        & 0.1014     &0.1964      & 0.2908    \\
          &FDR$_{\Psi}$   &   0.0724    &  0.1572          & 0.2485       & 0.0756     &0.1623      & 0.2539  \\
          &FDR$_{B}$      &   0.0369    &  0.1090          & 0.1944        & 0.0547     &0.1343     & 0.2225  \\
               &FDR$_{RB}$    &   0.0609   &  0.1433          & 0.2337        & 0.0722      &0.1599      & 0.2512  \\
           \hline
      & \multicolumn{6}{c}{Lognormal(0,1)}&\multicolumn{1}{c|}{} \\
      \hline
       500 & FDR$_{\Phi}$   &   0.0810    &  0.1693          & 0.2667        &0.0761     &0.1617      & 0.2560   \\
          &FDR$_{\Psi}$   &   0.0432    &  0.1123         & 0.1964      & 0.0519    &0.1297      & 0.2144  \\
          &FDR$_{B}$    &   0.0005    & 0.0103         & 0.0425        & 0.0059     &0.0384     & 0.0960  \\
               &FDR$_{RB}$    &   0.0300   &  0.0919          & 0.1697        & 0.0466      &0.1187      & 0.2086  \\
          \hline
      1000 & FDR$_{\Phi}$   &   0.0799    &  0.1701         & 0.2657        & 0.0760     &0.1628      & 0.2572   \\
          &FDR$_{\Psi}$    &   0.0433    &  0.1133          & 0.1962        & 0.0521     &0.1296      & 0.2165  \\
          &FDR$_{B}$  &   0.0004    & 0.0137          & 0.0504        & 0.0064     &0.0418      & 0.1032  \\
               &FDR$_{RB}$    &  0.0339   &  0.0953          & 0.1748        & 0.0485      &0.1237      & 0.2083  \\
          \hline
      3000 & FDR$_{\Phi}$  &   0.0805    &  0.1704         & 0.2654       & 0.0749     &0.1629      & 0.2578    \\
          &FDR$_{\Psi}$    &   0.0442   &  0.1142         & 0.1982        & 0.0523     &0.1283     & 0.2179  \\
          &FDR$_{B}$       &   0.0008    &  0.0151         & 0.0507       & 0.0070    &0.0432     & 0.1052  \\
               &FDR$_{RB}$    &   0.0319   &  0.0952          & 0.1766        & 0.0488      &0.1239      & 0.2129  \\
          \hline
    \end{tabular}

    \label{tb:simu3}
  \end{center}
\end{table}

\begin{table}[hptb]\small\addtolength{\tabcolsep}{-4pt}
  \begin{center}
    \caption{Comparison of power (FDR=$\alpha$)}
    \begin{tabular}{|c c| @{\hspace{2em}}c@{\hspace{1em}}c@{\hspace{1em}}c|@{\hspace{1em}}c@{\hspace{1em}}c@{\hspace{1em}} c| }
      \hline
      & \multicolumn{4}{c}{$n=30$}&\multicolumn{3}{c|}{$n=50$} \\
      \hline
  $m$  & $ |\quad { \alpha}$ &  0.1 &0.2& 0.3& 0.1&
      0.2 & 0.3\\
      \hline
      & \multicolumn{6}{c}{t(4)}&\multicolumn{1}{c|}{} \\
      \hline
       500 & power$_{\Phi}$    & 0.8305   &  0.8890          & 0.9190        & 0.8266     &0.8853     & 0.9173    \\
          &power$_{\Psi}$   &   0.7782    &  0.8576          & 0.9007        & 0.7968     &0.8684      & 0.9058  \\
          &power$_{B}$    &   0.6901    &  0.8190          & 0.8746       & 0.7582     &0.8574      & 0.8984  \\
               &power$_{RB}$    &   0.7554   &  0.8439          & 0.8916        & 0.7908      &0.8676      & 0.9072  \\
          \hline
      1000 & power$_{\Phi}$    &   0.8633   &  0.9113         & 0.9369        & 0.8648     &0.9144      & 0.9403    \\
          &power$_{\Psi}$  &   0.8200   &  0.8869         & 0.9208        & 0.8389    &0.8998      & 0.9315  \\
          &power$_{B}$    &   0.7472    & 0.8477          & 0.8977        & 0.8050     &0.8838      & 0.9219  \\
               &power$_{RB}$    &   0.8021   &  0.8788          & 0.9161        & 0.8357      &0.8992      & 0.9305  \\
          \hline
      3000 & power$_{\Phi}$  &   0.9078    &  0.9413          & 0.9589        & 0.9091     &0.9434      & 0.9605    \\
          &power$_{\Psi}$   &   0.8768    &  0.9249          & 0.9485        & 0.8915     &0.9339      & 0.9549  \\
          &power$_{B}$      &   0.8305    &  0.9053          & 0.9384        & 0.8755     &0.9293    & 0.9533  \\
               &power$_{RB}$    &   0.8651   &  0.9203          & 0.9455        & 0.8913      &0.9350      & 0.9555  \\
           \hline
      & \multicolumn{6}{c}{Lognormal(0,1)}&\multicolumn{1}{c|}{} \\
      \hline
       500 & power$_{\Phi}$   &  0.7916    &  0.8453          & 0.8796        &0.7789     &0.8390      & 0.8764   \\
          &power$_{\Psi}$   &   0.7424    &  0.8165         & 0.8561      & 0.7507    &0.8209      & 0.8615  \\
          &power$_{B}$    &   0.3216    &  0.6267         & 0.7404        & 0.5426     &0.7275     & 0.8037  \\
               &power$_{RB}$    &   0.7217   &  0.8044          & 0.8486        & 0.7479      &0.8203      & 0.8623  \\
          \hline
      1000 & power$_{\Phi}$   &   0.8240    &  0.8703         & 0.8989        & 0.8156     &0.8669      & 0.8975   \\
          &power$_{\Psi}$    &  0.7842    &  0.8444          & 0.8795        & 0.7899     &0.8506      & 0.8859  \\
          &power$_{B}$  &   0.4340    & 0.6978          & 0.7898        & 0.6320     &0.7749      & 0.8379  \\
               &power$_{RB}$    &   0.7647   &  0.8343          & 0.8715        & 0.7869     &0.8499      & 0.8859 \\
          \hline
      3000 & power$_{\Phi}$  &   0.8634    &  0.9003         & 0.9224        & 0.8610     &0.9021      & 0.9257    \\
          &power$_{\Psi}$    &   0.8314   &  0.8805         & 0.9079        & 0.8415     &0.8895     & 0.9169  \\
          &power$_{B}$       &   0.5880    &  0.7688         & 0.8386        & 0.7192    &0.8300      & 0.8780  \\
               &power$_{RB}$    &   0.8140   &  0.8711         & 0.9018       & 0.8374      &0.8865      & 0.9149  \\
          \hline
    \end{tabular}

    \label{tb:simu3}
  \end{center}
\end{table}

\section{Proof of Main Results}

By Theorem 1.2 in Wang (2005) and equation (2.2) in Shao (1999), we have for $0\leq t\leq o(n^{1/4})$,
\begin{eqnarray}\label{le7}
\pr(|T_{i}-\sqrt{n}\mu_{i}/\hat{s}_{n}|\geq t)=\frac{1}{2}G(t)\Big{[}\exp\Big{(}-\frac{t^{3}}{3\sqrt{n}}\kappa_{i}\Big{)}+
\exp\Big{(}\frac{t^{3}}{3\sqrt{n}}\kappa_{i}\Big{)}\Big{]}(1+o(1)),
\end{eqnarray}
where $o(1)$ is uniformly in $1\leq i\leq m$, $G(t)=2-2\Phi(t)$ and $\kappa_{i}=\ep Y_{i}^{3}$.

For any $b_{m}\rightarrow\infty$ and $b_{m}=o(m)$, we first prove that, under (C1$^{*}$) and $\log m=o(n^{1/2})$ (or (C1) and  $\log m=O(n^{\zeta})$ for some $0<\zeta<3/23$),
\begin{eqnarray}\label{aa133}
\sup_{0\leq t\leq G^{-1}_{\kappa}(b_{m}/m)}\Big{|}\frac{\sum_{i\in\mathcal{H}_{0}}I\{|T_{i}|\geq t\}}{m_{0}G_{\kappa}(t)}-1\Big{|}\rightarrow 0
\end{eqnarray}
in probability, where
\begin{eqnarray*}
G_{\kappa}(t)=\frac{1}{2m_{0}}G(t)\sum_{i\in\mathcal{H}_{0}}\Big{[}\exp\Big{(}-\frac{t^{3}}{3\sqrt{n}}\kappa_{i}\Big{)}+
\exp\Big{(}\frac{t^{3}}{3\sqrt{n}}\kappa_{i}\Big{)}\Big{]}=:G(t)\hat{\kappa}_{\Phi}(t)
\end{eqnarray*}
 and $G^{-1}_{\kappa}(t)=\inf\{y\geq 0: G_{\kappa}(y)=t\}$ for $0\leq t\leq 1$. Note that for $0\leq t\leq o(\sqrt{n})$, $G_{\kappa}(t)$ is a strictly decreasing and continuous function. Let $z_{0}<z_{1}<\cdots<z_{d_{m}}\leq 1$ and $t_{i}=G^{-1}_{\kappa}(z_{i})$, where $z_{0}=b_{m}/m$, $z_{i}=b_{m}/m+b^{2/3}_{m}e^{i^{\delta}}/m$,
$d_{m}=[\{\log ((m-b_{m})/b^{2/3}_{m})\}^{1/\delta}]$ and $0<\delta<1$ which will be specified later. Note that $G_{\kappa}(t_{i})/G_{\kappa}(t_{i+1})=1+o(1)$ uniformly in $i$, and
$t_{0}/\sqrt{2\log (m/b_{m})}=1+o(1)$.
Then, to prove (\ref{aa133}), it is enough to show that
\begin{eqnarray}\label{aa13}
\sup_{0\leq j\leq d_{m}}\Big{|}\frac{\sum_{i\in\mathcal{H}_{0}}I\{|T_{i}|\geq t_{j}\}}{m_{0}G_{\kappa}(t_{j})}-1\Big{|}\rightarrow 0
\end{eqnarray}
in probability.
Under (C1), define
\begin{eqnarray*}
\mathcal{S}_{j}=\{i\in\mathcal{H}_{0}: |r_{ij}|\geq (\log m)^{-1-\gamma}\},\quad \mathcal{S}_{j}^{c}=\mathcal{H}_{0}-\mathcal{S}_{j},
\end{eqnarray*}
and under (C1$^{*}$), define
\begin{eqnarray*}
\mathcal{S}_{j}=\{i\in\mathcal{H}_{0}: \mbox{$X_{i}$ is dependent with $X_{j}$}\}.
\end{eqnarray*}
We claim that, under (C1$^{*}$) and $\log m=o(n^{1/2})$ (or (C1) and  $\log m=O(n^{\zeta})$ for some $0<\zeta<3/23$),
for any $\varepsilon>0$ and some $\gamma_{1}>0$,
\begin{eqnarray}\label{a1}
I_{2}(t)&:=&\ep\Big{(}\sum_{i\in \mathcal{H}_{0}}\{I\{T_{i}\geq t\}-\pr(|T_{i}|\geq t)\}\Big{)}^{2}\cr
&\leq& Cm^{2}_{0}G^{2}_{\kappa}(t)\Big{(}\frac{1}{m_{0}G_{\kappa}(t)}+\frac{\exp\Big{(}(r+\varepsilon)t^{2}/(1+r)\Big{)}}{m^{1-\rho}}+(\log m)^{-1-\gamma_{1}}\Big{)}
\end{eqnarray}
uniformly in $t\in[0, K\sqrt{\log m}]$ for all $K>0$. Take $(1+\gamma_{1})^{-1}<\delta<1$. By (\ref{a1}) and $G^{-1}_{\kappa}(b_{m}/m)\sim \sqrt{2\log (m/b_{m})}$,
for any $\varepsilon>0$, we have
\begin{eqnarray*}
&&\sum_{j=0}^{d_{m}}\pr\Big{(}\Big{|}\frac{\sum_{i\in \mathcal{H}_{0}}I\{T_{i}\geq t_{j}\}}{m_{0}G_{\kappa}(t_{j})}-1\Big{|}\geq \varepsilon\Big{)}\cr
&&\leq \sum_{j=0}^{d_{m}}\pr\Big{(}\Big{|}\frac{\sum_{i\in \mathcal{H}_{0}}(I\{T_{i}\geq t_{j}\}-\pr(|T_{i}|\geq t_{j})}{m_{0}G_{\kappa}(t_{j})}\Big{|}\geq
\varepsilon/2\Big{)}\cr
&&\leq C\Big{(}\frac{1}{m_{0}G_{\kappa}(t_{0})}+\sum_{j=1}^{d_{m}}\frac{1}{m_{0}G_{\kappa}(t_{j})}+d_{m}m^{-1+\rho+\frac{2r+2\varepsilon}{1+r}+o(1)}
+d_{m}(\log m)^{-1-\gamma_{1}}\Big{)}\cr
&&\leq C\Big{(}b_{m}^{-1}+b_{m}^{-2/3}\sum_{j=1}^{d_{m}}e^{-j^{\delta}}+o(1)\Big{)}=o(1).
\end{eqnarray*}
This prove (\ref{aa13})

To prove (\ref{a1}), we need  the following lemma which will be proved  in the
supplementary file.

\begin{lemma}\label{le1} (i). Suppose that $\log m=O(n^{1/2})$. For any $\varepsilon>0$,
\begin{eqnarray}\label{a8}
\max_{j\in\mathcal{H}_{0}}\max_{i\in \mathcal{S}_{j}\setminus j}\pr\Big{(}|T_{i}|\geq t,|T_{j}|>t\Big{)}\leq C\exp(-(1-\varepsilon)t^{2}/(1+r))
\end{eqnarray}
uniformly in $t\in[0,o(n^{1/4}))$.

(ii). Suppose that $\log m=O(n^{\zeta})$ for some $0<\zeta<3/23$. We have for any $K>0$
\begin{eqnarray}\label{a9}
\pr\Big{(}|T_{i}|>t,|T_{j}|>t\Big{)}=(1+A_{n})\pr(|T_{i}|>t)\pr(|T_{j}|>t)
\end{eqnarray}
uniformly in $0\leq t\leq K\sqrt{\log m}$, $j\in\mathcal{H}_{0}$ and $i\in \mathcal{S}_{j}^{c}$, where
$
|A_{n}|\leq C(\log m)^{-1-\gamma_{1}}
$
for some $\gamma_{1}>0$.
\end{lemma}

Set $f_{ij}(t)=\pr\Big{(}|T_{i}|\geq t,|T_{j}|\geq t\Big{)}-\pr\Big{(}|T_{i}|\geq t)\pr\Big{(}|T_{j}|\geq t\Big{)}$. Note that under (C1$^{*}$) $f_{ij}=0$ when $j\in\mathcal{H}_{0}\backslash\mathcal{S}_{i}$. We have
\begin{eqnarray*}
I_{2}(t)&\leq& \sum_{i\in\mathcal{H}_{0}}\sum_{j\in\mathcal{S}_{i}}\pr\Big{(}|T_{i}|\geq t,|T_{j}|\geq t\Big{)}
+\sum_{i\in\mathcal{H}_{0}}\sum_{j\in\mathcal{H}_{0}\backslash\mathcal{S}_{i}}f_{ij}(t)\cr
&\leq& Cm_{0}G_{\kappa}(t)+C\frac{\exp\Big{(}(r+2\varepsilon)t^{2}/(1+r)\Big{)}}{m^{1-\rho}}m^{2}_{0}G^{2}_{\kappa}(t)+A_{n}m^{2}_{0}G^{2}_{\kappa}(t),
\end{eqnarray*}
where the last inequality follows from Lemma 6.1 and $G_{\kappa}(t)=G(t)e^{o(1)t^{2}}$ for $t=o(\sqrt{n})$.   This proves (\ref{a1}). \qed

\subsection{Proof of Theorem \ref{th1} and Corollary \ref{co1}}

We only prove the theorem for $\hat{p}_{i,\Phi}$. The proof for $\hat{p}_{i,\Psi}$ is exactly the same by replacing $G(t)$ with
$2-2\Psi(t)$.
By Lemma 1 in Storey, Taylor and Siegmund (2004), we can see that  B-H method with $\hat{p}_{i,\Phi}$ is equivalent to the following procedure: reject $H_{0i}$ if and only if $\hat{p}_{i,\Phi}\leq \hat{t}_{0}$, where
\begin{eqnarray*}
\hat{t}_{0}=\sup\Big{\{}0\leq t\leq 1:~ t\leq \frac{\alpha\max(\sum_{1\leq i\leq m}I\{\hat{p}_{i,\Phi}\leq t\},1)}{m}\Big{\}}.
\end{eqnarray*}
It is equivalent to reject $H_{0i}$ if and only if $|T_{i}|\geq \hat{t}$, where
\begin{eqnarray*}
\hat{t}=\inf\Big{\{}t\geq 0:~ 2-2\Phi(t)\leq \frac{\alpha\max(\sum_{1\leq i\leq m}I\{|T_{i}|\geq t\},1)}{m}\Big{\}}.
\end{eqnarray*}
By the continuity of $\Phi(t)$ and the monotonicity of the indicator function, it is easy to see that
\begin{eqnarray*}
\frac{mG(\hat{t})}{\max(\sum_{1\leq i\leq m}I\{|T_{i}|\geq \hat{t}\},1)}=\alpha,
\end{eqnarray*}
where $G(t)=2-2\Phi(t)$. Let $\mathcal{M}$ be a subset of $\{1,2,\ldots,m\}$ satisfying $\mathcal{M}\subset \Big{\{}i: |\mu_{i}/\sigma_{i}|\geq 4\sqrt{\log m/n}\Big{\}}$ and Card$(\mathcal{M})\leq \sqrt{n}$.
By $\max_{1\leq i\leq m}\ep Y_{i}^{4}\leq K$ and Markov's inequality, for any $\varepsilon>0$,
\begin{eqnarray*}
\pr(\max_{i\in\mathcal{M}}|\hat{s}^{2}_{ni}/\sigma_{i}^{2}-1|\geq \varepsilon)=O(1/\sqrt{n}).
\end{eqnarray*}
This, together with  (\ref{c1}) and (\ref{le7}), implies that there exist some $c>\sqrt{2}$ and some $b_{m}\rightarrow\infty$,
\begin{eqnarray}\label{tv2}
\pr\Big{(}\sum_{i=1}^{m}I\{|T_{i}|\geq c\sqrt{\log m}\}\geq b_{m}\Big{)}\rightarrow 1.
\end{eqnarray}
This implies that
$
\pr\Big{(}\hat{t}\leq G^{-1}(\alpha b_{m}/m)\Big{)}\rightarrow 1
$ and $\pr(\hat{m}\geq b_{m})\rightarrow 1$.
By (\ref{aa133}) and $G_{\kappa}(t)\geq G(t)$, it follows that $\pr(\hat{t}\leq G_{\kappa}^{-1}(\alpha b_{m}/m))\rightarrow 1$. Therefore, by (\ref{aa133})
\begin{eqnarray*}
\frac{\sum_{i\in\mathcal{H}_{0}}I\{|T_{i}|\geq \hat{t}\}}{m_{0}G_{\kappa}(\hat{t})}\rightarrow 1
\end{eqnarray*}
in probability.
 Note that
\begin{eqnarray*}
G(\hat{t})=\frac{\alpha\hat{m}}{m}+\frac{\alpha m_{0}}{m}\frac{\sum_{i\in\mathcal{H}_{0}}I\{|T_{i}|\geq \hat{t}\}}{m_{0}},
\end{eqnarray*}
where $\hat{m}=\sum_{i\in\mathcal{H}_{1}}I\{|T_{i}|\geq \hat{t}\}$. With probability tending to one,
\begin{eqnarray}\label{prf3}
G(\hat{t})=\frac{\alpha\hat{m}}{m}+\frac{\alpha m_{0}}{m}G(\hat{t})\hat{\kappa}_{\Phi}(1+o(1))\geq \frac{\alpha m_{0}}{m}G(\hat{t})\hat{\kappa}_{\Phi}(1+o(1)).
\end{eqnarray}
So $\pr(\hat{\kappa}_{\Phi}\leq m/(\alpha m_{0})+\varepsilon)\rightarrow 1$ for any $\varepsilon>0$.  Let $\hat{\kappa}^{*}_{\Phi}=\hat{\kappa}_{\Phi}I\{\hat{\kappa}_{\Phi}\leq 2(\alpha(1-\gamma))^{-1})\}$. Note that $m/(\alpha m_{0})+\varepsilon\leq 2(\alpha(1-\gamma))^{-1}$.
We have
\begin{eqnarray*}
\frac{FDP_{\Phi}}{\frac{m_{0}}{m}\alpha \hat{\kappa}_{\Phi}^{*}}=\frac{\sum_{i\in\mathcal{H}_{0}}I\{|T_{i}|\geq \hat{t}\}}{m_{0}G_{\kappa}(\hat{t})}\frac{\hat{\kappa}_{\Phi}}{\hat{\kappa}_{\Phi}^{*}}(1+o(1))\rightarrow 1
\end{eqnarray*}
in probability, where $FDP_{\Phi}$ is the false discovery proportion $\text{V}/(\text{R}\vee 1)$.
Then for any $\varepsilon>0$,
\begin{eqnarray*}
FDR_{\Phi}\leq (1+\varepsilon)\frac{m_{0}}{m}\alpha\ep  \hat{\kappa}_{\Phi}^{*}+\pr\Big{(}FDP_{\Phi}\geq (1+\varepsilon)\frac{m_{0}}{m}\alpha \hat{\kappa}_{\Phi}^{*}\Big{)}
\end{eqnarray*}
and
\begin{eqnarray*}
FDR_{\Phi}\geq (1-\varepsilon)\frac{m_{0}}{m}\alpha\ep  \hat{\kappa}_{\Phi}^{*}-2(\alpha(1-\gamma))^{-1}\pr\Big{(}FDP_{\Phi}\leq (1-\varepsilon)\frac{m_{0}}{m}\alpha \hat{\kappa}_{\Phi}^{*}\Big{)}.
\end{eqnarray*}
This proves the Theorem 2.1. Corollary \ref{co1} (1) follows directly from Theorem 2.1 and $\pr(\hat{t}\leq \sqrt{2\log m})\rightarrow 1$.

To prove Corollary \ref{co1} (2), we first assume that $\frac{\alpha m_{0}}{m}\hat{\kappa}_{\Phi}\leq 1-\eta$ for some $(1-\eta)/\alpha>1$. So, by (\ref{prf3}) and the condition $m_{1}=\exp(o(n^{1/3}))$, with probability tending to one,
$G(\hat{t})\leq 2\alpha\eta^{-1}\hat{m}/m\leq 2\alpha\eta^{-1}m^{-1+o(1)}$. Hence $\hat{t}\geq c\sqrt{\log m}$ for any $c<\sqrt{2}$. Recall that
$\tau=\liminf_{m\rightarrow\infty}m_{0}^{-1}\sum_{i\in\mathcal{H}_{0}}|\ep Y_{i}^{3}|>0$. Set
\begin{eqnarray*}
\mathcal{H}_{01}=\{i\in\mathcal{H}_{0}: |\ep Y_{i}^{3}|\geq \tau/8\}.
\end{eqnarray*}
By the definition of $\tau$ and $|\ep Y_{i}^{3}|\leq (\ep (Y_{i}^{4})^{3/4}\leq b_{0}^{3/4}$,
$
m_{0}^{-1}|\mathcal{H}^{c}_{01}|\tau/8+b_{0}^{3/4}m_{0}^{-1}|\mathcal{H}_{01}|\geq \tau/2.
$
This implies that $|\mathcal{H}_{01}|\geq \tau b_{0}^{-3/4}m_{0}/4$. Hence we can get
$m_{0}^{-1}\sum_{i\in\mathcal{H}_{0}}|\ep Y_{i}^{3}|^{2}\geq c_{\tau}$ for some $c_{\tau}>0$. It follows from Taylor's expansion of the exponential function and $\hat{t}\geq c\sqrt{\log m}$ that $\hat{\kappa}_{\Phi}\geq 1+\epsilon$ for some $\epsilon>0$. On the other hand, if $\frac{\alpha m_{0}}{m}\hat{\kappa}_{\Phi}> 1-\eta$, then
$\hat{\kappa}_{\Phi}\geq 1+\epsilon$ for some $\epsilon>0$. This yields that
$\pr(\hat{\kappa}_{\Phi}\geq 1+\epsilon)\rightarrow 1$ for some $\epsilon>0$. So we have $\kappa_{\Phi}\geq 1+\epsilon$ for some $\epsilon>0$.
Note that $m_{0}/m\rightarrow 1$. We prove Corollary \ref{co1} (2).

We next prove Corollary \ref{co1} (3). By the inequality $e^{x}+e^{-x}\geq |x|$, $\pr(\hat{\kappa}_{\Phi}\leq m/(\alpha m_{0})+\varepsilon)\rightarrow 1$,
we obtain that
\begin{eqnarray*}
\frac{\sum_{i\in\mathcal{H}_{0}}\frac{\hat{t}^{3}}{\sqrt{n}}|\ep Y_{i}^{3}|}{2m_{0}}\leq  m/(\alpha m_{0})+\varepsilon
\end{eqnarray*}
with probability tending to one. By $\tau>0$, we have
 $\pr(\hat{t}\leq cn^{1/6})\rightarrow 1$  for some constant $c>0$. So  $\pr(G(\hat{t})\geq \exp(-2cn^{1/3})\rightarrow 1$. Since $\hat{m}/m\leq \exp(-Mn^{1/3})$ for any $M>0$, we have by (\ref{prf3})
\begin{eqnarray*}
\frac{\alpha m_{0}}{m}\hat{\kappa}_{\Phi}\rightarrow 1.
\end{eqnarray*}
in probability.
Hence $\kappa_{\Phi}\rightarrow 1/\alpha$ since $m_{0}/m\rightarrow 1$. The proof is finished.\qed

\subsection{Proof of Theorems \ref{th2-2} and \ref{th22}}

Let  $\hat{\kappa}_{i}=\frac{1}{n\hat{s}_{ni}^{3}}\sum_{k=1}^{n}(X_{ki}-\bar{X}_{i})^{3}$.
Define the event
\begin{eqnarray*}
\F=\{\max_{1\leq i\leq m}\frac{1}{n\hat{s}_{ni}^{4}}\sum_{k=1}^{n}(X_{ki}-\bar{X}_{i})^{4}\leq K_{1},\max_{1\leq i\leq m}|\hat{\kappa}_{i}-\kappa_{i}|\leq K_{2}\sqrt{\log m/n}\}
\end{eqnarray*}
 for some large $K_{1}>0$ and $K_{2}>0$. We first suppose that $\pr(\F)\rightarrow 1$.
  Let $G^{*}_{i}(t)=\pr^{*}(|T^{*}_{ki}|\geq t)$ be the conditional distribution of $T^{*}_{ki}$
given $\mathcal{X}=\{\mathcal{X}_{1},\cdots,\mathcal{X}_{m}\}$.  Note that, given $\mathcal{X}$ and on the event $\F$,
\begin{eqnarray*}
G^{*}_{i}(t)&=&\frac{1}{2}G(t)\Big{[}\exp\Big{(}-\frac{t^{3}}{3\sqrt{n}}\hat{\kappa}_{i}\Big{)}+
\exp\Big{(}\frac{t^{3}}{3\sqrt{n}}\hat{\kappa}_{i}\Big{)}\Big{]}(1+o(1))\cr
&=&\frac{1}{2}G(t)\Big{[}\exp\Big{(}-\frac{t^{3}}{3\sqrt{n}}\kappa_{i}\Big{)}+
\exp\Big{(}\frac{t^{3}}{3\sqrt{n}}\kappa_{i}\Big{)}\Big{]}(1+o(1))
\end{eqnarray*}
uniformly in $0\leq t\leq o(n^{1/4})$. Hence,  given $\mathcal{X}$ and on the event $\F$,
\begin{eqnarray}\label{bap}
\frac{G^{*}_{i}(t)}{\pr(|T_{i}-\sqrt{n}\mu_{i}/\hat{s}_{n}|\geq t)}=1+o(1)
\end{eqnarray}
uniformly in $1\leq i\leq m$ and $0\leq t\leq o(n^{1/4})$.
 Put
\begin{eqnarray*}
\hat{G}_{\kappa}(t)=\frac{1}{2m}G(t)\sum_{1\leq i\leq m}\Big{[}\exp\Big{(}-\frac{t^{3}}{3\sqrt{n}}\kappa_{i}\Big{)}+
\exp\Big{(}\frac{t^{3}}{3\sqrt{n}}\kappa_{i}\Big{)}\Big{]}.
\end{eqnarray*}
Set $\hat{c}_{m}=\hat{G}^{-1}_{\kappa}(b_{m}/m)$. Note that, given $\mathcal{X}$, $T_{ki}^{*}$, $1\leq k\leq N$, $1\leq i\leq m$, are independent. Hence, as (\ref{aa133}), we can show that for any $b_{m}\rightarrow\infty$,
\begin{eqnarray}\label{aa14}
\sup_{0\leq t\leq \hat{c}_{m}}\Big{|}\frac{G^{*}_{N,m}(t)}{\hat{G}_{\kappa}(t)}-1\Big{|}\rightarrow 0
\end{eqnarray}
in probability.  For $t=O(\sqrt{\log m})$, under the conditions of Theorem 3.2, we have $\hat{G}_{\kappa}(t)/G_{\kappa}(t)=1+o(1)$. So, it is easy to see that (\ref{aa133}) still holds when
 $G^{-1}_{\kappa}(b_{m}/m)$ is replaced by $\hat{G}^{-1}_{\kappa}(b_{m}/m)$.
This  implies that for any $b_{m}\rightarrow\infty$,
\begin{eqnarray}\label{aa16}
\sup_{0\leq t\leq \hat{c}_{m}}\Big{|}\frac{\sum_{i\in\mathcal{H}_{0}}I\{|T_{i}|\geq t\}}{m_{0}G^{*}_{N,m}(t)}-1\Big{|}\rightarrow 0
\end{eqnarray}
in probability.

Let
\begin{eqnarray*}
\hat{t}_{0}=\sup\Big{\{}0\leq t\leq 1:~ t\leq \frac{\alpha\max(\sum_{1\leq i\leq m}I\{\hat{p}_{i,B}\leq t\},1)}{m}\Big{\}}.
\end{eqnarray*}
Then we have
\begin{eqnarray*}
\hat{t}_{0}=\frac{\alpha\max(\sum_{1\leq i\leq m}I\{\hat{p}_{i,B}\leq \hat{t}_{0}\},1)}{m}.
\end{eqnarray*}
By (\ref{le7}) and (\ref{bap}), we have, given $\mathcal{X}$ and on the event $\F$, $G^{*}_{i}(c\sqrt{\log m})=m^{-c^{2}/2+o(1)}$ for any $c>\sqrt{2}$ uniformly in $i$.
So, by Markov's inequality, for any $\varepsilon>0$, we have $\pr\Big{(}G^{*}_{N,m}(c\sqrt{\log m})\leq m^{-c^{2}/2+\varepsilon}\Big{)}\rightarrow 1$.
By (\ref{c1}) and (\ref{tv2}), we have $\pr(\hat{t}_{0}\geq \alpha b_{m}/m)\rightarrow 1$ for some $b_{m}\rightarrow\infty$. It follows from (\ref{aa16}) that
\begin{eqnarray*}
\frac{\sum_{i\in\mathcal{H}_{0}}I\{\hat{p}_{i,B}\leq \hat{t}_{0}\}}{m_{0}\hat{t}_{0}}\rightarrow 1
\end{eqnarray*}
in probability. This finishes the proof of Theorem \ref{th2-2} if we can show that $\pr(\F)\rightarrow 1$. Without loss of generality, we can assume that
$\mu_{i}=0$ and $\sigma_{i}=1$.
We first show that for some constant $K_{1}>0$,
\begin{eqnarray}\label{aa17}
\pr\Big{(}\max_{1\leq i\leq m}\Big{|}\sum_{k=1}^{n}(X_{ki}^{4}-\ep X_{ki}^{4})\Big{|}\geq K_{1}n\Big{)}=o(1).
\end{eqnarray}
For $1\leq i\leq n$, put
 \begin{eqnarray*}
 \hat{X}_{ki}=X_{ki}I\{|X_{ki}|\leq \sqrt{n/\log m}\},\quad \breve{X}_{ki}=X_{ki}- \hat{X}_{ki}.
 \end{eqnarray*}
Then, for large $n$,
\begin{eqnarray*}
&&\pr\Big{(}\max_{1\leq i\leq m}\Big{|}\sum_{k=1}^{n}(\breve{X}_{ki}^{4}-\ep \breve{X}_{ki}^{4})\Big{|}\geq K_{1}n/2\Big{)}\cr
&&\quad\leq nm\max_{1\leq i\leq m}\pr(|X_{1i}|\geq \sqrt{n/\log m})\cr
&&\quad\leq C\exp(\log m+\log n-tn/\log m)\cr
&&\quad=o(1).
\end{eqnarray*}
Let $Z_{ki}=\hat{X}_{ki}^{4}-\ep \hat{X}_{ki}^{4}$. By the inequality
$|e^{s}-1-s|\leq s^{2}e^{\max(s,0)}$ and $1+s\leq e^{s}$, we have for $\eta=2^{-1}t(\log m)/n$ and some large $K_{1}$
\begin{eqnarray*}
&&\pr\Big{(}\max_{1\leq i\leq m}\Big{|}\sum_{k=1}^{n}Z_{ki}\Big{|}\geq K_{1}n/2\Big{)}\cr
&&\quad\leq \sum_{i=1}^{m}\pr\Big{(}\sum_{k=1}^{n}Z_{ki}\geq K_{1}n/2\Big{)}+ \sum_{i=1}^{m}\pr\Big{(}-\sum_{k=1}^{n}Z_{ki}\geq K_{1}n/2\Big{)}\cr
&&\quad\leq \sum_{i=1}^{m}\exp(-\eta K_{1}n/2)\Big{[}\prod_{k=1}^{n}\exp(\eta Z_{ki})+\prod_{k=1}^{n}\exp(-\eta Z_{ki})\Big{]}\cr
&&\quad\leq 2\sum_{i=1}^{m}\exp(-\eta K_{1}n/2+\eta^{2}n\ep Z_{1i}^{2}e^{\eta|Z_{1i}|})\cr
&&\quad\leq C\exp(\log m-t K_{1}(\log m)/4)\cr
&&\quad=o(1).
\end{eqnarray*}
This proves (\ref{aa17}). By replacing $X_{ki}^{4}$, $\eta=2^{-1}t(\log m)/n$ and $K_{1}n/2$ with $X_{ki}^{3}$,
$\eta=2^{-1}t\sqrt{(\log m)/n}$ and $K_{1}\sqrt{n\log m}/2$ respectively in the above proof, we can show that
\begin{eqnarray}\label{aa18}
\pr\Big{(}\max_{1\leq i\leq m}\Big{|}\frac{1}{n}\sum_{k=1}^{n}(X_{ki}^{3}-\ep X_{ki}^{3})\Big{|}\geq K_{1}\sqrt{(\log m)/n}\Big{)}=o(1).
\end{eqnarray}
Similarly, we have
\begin{eqnarray}\label{aa19}
\pr\Big{(}\max_{1\leq i\leq m}\Big{|}\frac{1}{n}\sum_{k=1}^{n}(X_{ki}^{2}-\ep X_{ki}^{2})\Big{|}\geq K_{1}\sqrt{(\log m)/n}\Big{)}=o(1)
\end{eqnarray}
and
\begin{eqnarray}\label{aa199}
\pr\Big{(}\max_{1\leq i\leq m}\Big{|}\frac{1}{n}\sum_{k=1}^{n}(X_{ki}-\ep X_{ki})\Big{|}\geq K_{1}\sqrt{(\log m)/n}\Big{)}=o(1).
\end{eqnarray}
Combining (\ref{aa17})-(\ref{aa199}), we prove that $\pr(\F)\rightarrow 1$.
\qed

\subsection{Proof of Theorems \ref{th2-222} and \ref{th222}}

Let
\begin{eqnarray*}
\hat{\F}=\{\max_{1\leq i\leq m}\frac{1}{n\hat{\sigma}_{i}^{4}}\sum_{k=1}^{n}(\hat{X}_{ki}-\hat{X}_{i})^{4}\leq K_{1},\max_{1\leq i\leq m}|\hat{\kappa}_{i}(\lambda_{ni})-\kappa_{i}|\leq K_{2}\sqrt{\log m/n}\}
\end{eqnarray*}
By the proof of Theorems \ref{th2-2} and \ref{th22}, it is enough to show $\pr(\hat{\F})\rightarrow 1$. Recall that $\hat{X}_{ki}=X_{ki}I\{|X_{ki}|\leq \lambda_{ni}\}$ and put
$Z_{ki}=\hat{X}^{4}_{ki}-\ep \hat{X}^{4}_{ki}$. Take $\eta=(\log m)/n$. We have
\begin{eqnarray*}
&&\pr\Big{(}\max_{1\leq i\leq m}\Big{|}\sum_{k=1}^{n}Z_{ki}\Big{|}\geq K_{1}n/2\Big{)}\cr
&&\quad\leq 2\sum_{i=1}^{m}\exp(-\eta K_{1}n/2+\eta^{2}n\ep Z_{1i}^{2}e^{\eta|Z_{1i}|})\cr
&&\quad\leq C\exp(2\log m-K_{1}(\log m)/4)\cr
&&\quad=o(1).
\end{eqnarray*}
Similarly, by replacing $\hat{X}_{ki}^{4}$, $\eta=(\log m)/n$ and $K_{1}n/2$ with $\hat{X}_{ki}^{3}$,
$\eta=\sqrt{(\log m)/n}$ and $K_{1}\sqrt{n\log m}/2$ respectively in the above proof, we can show that
\begin{eqnarray*}
\pr\Big{(}\max_{1\leq i\leq m}\Big{|}\frac{1}{n}\sum_{k=1}^{n}(\hat{X}_{ki}^{3}-\ep \hat{X}_{ki}^{3})\Big{|}\geq K_{1}\sqrt{(\log m)/n}\Big{)}=o(1).
\end{eqnarray*}
Also, using the above arguments, it is easy to show that
\begin{eqnarray*}
\pr\Big{(}\max_{1\leq i\leq m}\Big{|}\frac{1}{n}\sum_{k=1}^{n}(\hat{X}_{ki}^{2}-\ep \hat{X}_{ki}^{2})\Big{|}\geq K_{1}\sqrt{(\log m)/n}\Big{)}=o(1)
\end{eqnarray*}
and
\begin{eqnarray*}
\pr\Big{(}\max_{1\leq i\leq m}\Big{|}\frac{1}{n}\sum_{k=1}^{n}(\hat{X}_{ki}-\ep \hat{X}_{ki})\Big{|}\geq K_{1}\sqrt{(\log m)/n}\Big{)}=o(1).
\end{eqnarray*}
Note that
\begin{eqnarray*}
\max_{1\leq i\leq m}\ep |X_{1i}|^{3}I\{|X_{1i}|\geq \lambda_{ni}\}
\leq C\sqrt{\frac{\log m}{n}}\max_{1\leq i\leq m}\ep X_{1i}^{6}
\end{eqnarray*}
and
\begin{eqnarray*}
\max_{1\leq i\leq m}\ep |X_{1i}|^{2}I\{|X_{1i}|\geq \lambda_{ni}\}\leq C\Big{(}\frac{\log m}{n}\Big{)}^{2/3}\max_{1\leq i\leq m}\ep X_{1i}^{6}.
\end{eqnarray*}
This proves $\pr(\hat{\F})\rightarrow 1$.\qed

\subsection{Proof of Theorem \ref{th21}}

Recall that
\begin{eqnarray*}
\frac{mG(\hat{t})}{\max(\sum_{1\leq i\leq m}I\{|T_{i}|\geq \hat{t}\},1)}=\alpha.
\end{eqnarray*}
From (\ref{tv2}), we have $
\pr\Big{(}\hat{t}\geq G^{-1}(\alpha b_{m}/m)\Big{)}\rightarrow 1.
$
The theorem follows from (\ref{aa133}) and the fact $G_{\kappa}(t)/G(t)=1+o(1)$ uniformly in $t\in[0,o(n^{1/6}))$. \qed

\end{document}